\documentclass[12pt]{article}
\usepackage[dvips]{graphicx}
\begin{document}

\begin{center}   {Dark Matter in the Local Group of Galaxies}
\end{center}
\vspace{.2in}
\begin{center}
P.D. Morley \\  
System of Systems Analytics, Inc. \\ 11250 Waples Mill Road \\ Suite 300 \\Fairfax, VA 22030-7400 \\
{\it E-mail address:} peter3@uchicago.edu \\
\vspace{0.25in}
and \\
\vspace{0.25in}
 D. J. Buettner \\
Aerospace Corporation \\P. O. Box 92957 \\Los Angeles, CA 90009-2957 \\
{\it E-mail address:}  Douglas.J.Buettner@aero.org
\end{center}

\begin{abstract}
We describe the neutrino flavor (e = electron, $\mu$ = muon, $\tau$ = tau) masses as $m_{i = e, \mu, \tau} = m + \Delta m_{i}$ with $\frac{|\Delta m_{i}|}{m} < 1$ and probably $\frac{|\Delta m_{i}|}{m} \ll 1$. The quantity $m$ is the degenerate neutrino mass. Because neutrino flavor is not a quantum number, this degenerate mass appears in the neutrino equation of state\cite{Mor1}. We apply a Monte Carlo computational physics technique to the Local Group (LG) of galaxies to determine an approximate location for a Dark Matter embedding condensed neutrino object\cite{Mor2} (CNO). The calculation is based on the rotational properties of the only spiral galaxies within the LG: M31, M33 and the Milky Way. CNOs could be the Dark Matter everyone is looking for and we estimate the CNO embedding the LG to have a mass  5.17$\times10^{15}$  M$_{\odot}$ and a  radius 1.316 Mpc, with the estimated value of $m \simeq 0.8$ eV/c$^{2}$. The up-coming KATRIN experiment \cite{KA} will either be the definitive result or eliminate condensed neutrinos as a Dark Matter candidate.
\end{abstract}


PACS Number(s): 95.35.+d, 95.30.cq, 98.35.Gi, 98.80.cq
\newpage

\tableofcontents
\section{Introduction}
The rotational properties of spiral galaxies are the original experimental evidence\cite{ZW} that Dark Matter must exist. Subsequently, modern cosmology in the form of Cold Dark Matter\footnote{A good history can be found online\cite{Hi}.} requires Dark Matter to be a large contributor to the energy density of the universe. If neutrinos are Dirac-type (not Majorana), they must posses a magnetic dipole moment, and it was shown\cite{Mor3} that the instantaneous relativistic radiation loss for a magnetic moment ${\cal P}_{M}$ is 
\begin{equation}
{\cal P}_{M} = \frac{2}{3 c^{3}}[\gamma^{8}(\vec{\beta} \cdot \dot{\vec{\beta}})^{2} \dot{\vec{\mu}^{2}} + \ldots ]
\end{equation}
where $\gamma^{-1} = \sqrt{1-\beta^{2}}$. This is a higher power of $\gamma$ than the  familiar Li\'{e}nard formula for charged particles Q
\begin{equation}
{\cal P}_{Q} = \frac{2}{3}\frac{Q^{2}}{c} \gamma^{6} [ (\dot{\vec{\beta}})^{2} -(\vec{\beta} \times \dot{\vec{\beta}})^{2} ] \; .
\end{equation}
Only in turbulent\footnote{Turbulent here means that the magnetic moment is flipping about.}, chaotic plasmas is this dipole radiation loss realizable. However, in the early universe, large, chaotic, and turbulent fields are expected\cite{Gra, Wag}. If the neutrinos lose energy in this manner, they are anticipated to condense and form degenerate neutrino matter. If this happens, the CNO are the largest and most massive objects in the universe\cite{Mor2} and become the `Dark Matter', which everybody is looking for. In this paper, we take the three spiral galaxies of the Local Group and use their rotational data to elucidate the properties of the CNO which appears to embed them.

\section{The spiral galaxies of the Local Group (LG)}
M33, M31 and the Milky Way are the only spiral galaxies of the Local group. The rotational data of M33\cite{EC} and M31\cite{BN} possess a unique property: they have no noticable azimuth angle dependence\footnote{This means that if the rotation curve were folded over itself along a diameter, the absolute value of the rotation (speeds) are the same.}. The physics implication is that for a given fixed radial distance from the galaxy's center, the 360 degree rotation circle is a equipotential surface. For spherical Dark Matter, such as a CNO, this can only happen if the spin axis of the galaxy is aligned radially in the embedding Dark Matter sphere. Having two spiral galaxies with no noticable azimuth angle dependence in their rotation curves allows us to determine the center of the CNO by extrapolating the two spin axis back to a point, Figure 1. In practice, they will not actually cross, but join within an error volume that will be calculable. The Milky Way has azimuth angle dependence in its rotation curve, because of the wide variation\cite{OMW} of the rotational speed with fixed radial distance. This means that the Milky Way spin axis is `canted' in the CNO spherical symmetry.

\subsection{Extrapolation of the radially aligned spiral galaxy spin unit vector $\hat{L}$}

In Table 1, we list the J2000 equatorial coordinates\footnote{Wolfram Mathematica 11.0\cite{Al}.} of M33 and M31. To find the center of the CNO embedding the LG, we extrapolate the M31 and M33 spin axis using a Monte Carlo algorithm which is described in the Appendix. The spin axis of a spiral galaxy is described in reference\cite{spin}. The spin axis unit-vector $\hat{L}$ of M31 and M33 is given by
\begin{equation}
\hat{L} = L_{r}\hat{r} + L_{\theta}\hat{\theta} +L_{\phi}\hat{\phi}
\end{equation}
with
\begin{eqnarray}
\hat{r} & = & \cos \theta \cos \phi \hat{x} + \cos \theta \sin \phi \hat{y} + \sin \theta \hat{z} \nonumber \\
\hat{\theta}  & = & \sin \theta \cos \phi \hat{x} + \sin \theta \sin \phi \hat{y} - \cos \theta \hat{z} \nonumber \\
\hat{\phi} & = & -\sin \phi \hat{x} + \cos \phi \hat{y} 
\end{eqnarray}
where $\theta$ is the declination angle and $\phi$ is the right ascension angle of the galaxy in question. The physical components are given by observables $L_{r} = \Omega$, $L_{\phi} = Q$ and $L_{\theta} = CQ$ , where $\Omega$ is the axial ratio and $C = \tan (\pi - \alpha)$, where $\alpha $ is the position angle, with $Q = \sqrt{\frac{1 - L_{r}^{2}}{1+C^{2}}}$. We use a computer algorithm (see Appendix) that finds the smallest extrapolated M33 and M31 spin-axis intersection for values of $\Omega$ and $\alpha$ within $\pm$9 \% of their quoted values within the literature. This will reveal the center of the CNO to within a computable error.

\begin{table}
    \begin{tabular}{cccc}
\multicolumn{4}{l}{} \\ \hline
Name & Distance (kpc)  & Dec (degree)  & RA (degree) \\ \hline
M31 & 788.333 & 41.2689 & 10.6846 \\
M33 & 862.417 & 30.6581 & 23.4662 \\
Milky Way & 7.61113 & -29.0078 & 266.417
\end{tabular}
\caption{Coordinates of the LG spiral galaxies used in this paper.}
\end{table}

\begin{table}
    \begin{tabular}{cccc}
\multicolumn{4}{l}{} \\ \hline
Quantity & Derived value & Quoted value & Reference \\ \hline
M33 axial ratio & 0.653903 & 0.615  & reference\cite{OO} \\
M33 position angle & 22.1965$^{\circ}$ & $22^{\circ}$ & reference\cite{OP} \\
M31 axial ratio & 0.575201 &  .6  & reference\cite{AA2} \\
M31 positional angle & 40.1856$^{\circ}$ & 38$^{\circ}$ & reference\cite{ZZ} 
\end{tabular}
\caption{Monte Carlo derived Local Group Parameters}
\end{table}

In Table 2, we present the Monte Carlo's algorithm's estimate of the M33, M31 axial ratio and position angles that gives a small error in their extrapolated intersection.

\begin{figure}
\includegraphics[scale=0.35]{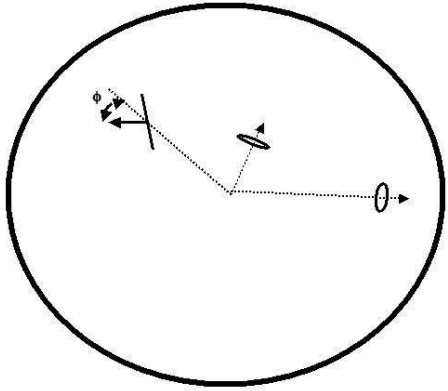}
\caption{Spiral galaxies in a CNO. Two galaxies have spins radially aligned and the third has a canted angle $\phi$.}
\label{Figure 1}
\end{figure}

\subsection{CNO parameters}
This spin axis extrapolation will reveal the approximate center of the CNO which is given in Table 3; there, the CNO is designated CNO-LG. The size and mass of the CNO are determined by three parameters: the degenerate neutrino mass $m$ and two boundary conditions\cite{Mor1,Mor2} at the center, of which the only one that varies is $x(0)$ where $x(0)$ is the quantity $\frac{P_{F}}{mc}$ at the center, with $P_{F}$ the Fermi momentum and $c$ the speed of light. In general, $x(0) \ll 1$ from fitting galaxy clusters embedded in CNO\cite{Mor2}. M33 has the best rotational data\cite{EC} because that reference includes an estimate for the contribution of M33's own mass. At about 15 kpc radial distance, the measured $\sqrt{<v^{2}>|_{R_{C}}} \sim$ 100 km/s, of which $\sim$ 40 - 50 km/s is the Dark Matter contribution. We now explain how this identifies the parameters of the CNO that embeds the Local Group.

\begin{table}
    \begin{tabular}{ccc}
\multicolumn{3}{l}{} \\ \hline
Quantity & Predicted value & unit \\ \hline
center CNO-LG distance from earth & 675.933 & kpc \\
center CNO-LG distance from M33 & 740.423 & kpc \\
center CNO-LG distance from M31 & 656.015 & kpc \\
algorithm error center CNO-LG & .706162e-3 & kpc \\
right ascension of CNO-LG center & -26.6588& $^{\circ}$ \\
declination of CNO-LG center & 0.91773 & $^{\circ}$ \\
galactic longitude of CNO-LG center & 62.83928785 & $^{\circ}$ \\
galactic latitude of CNO-LG center & -42.77834848 & $^{\circ}$ \\
Milky Way cant angle & 47.221 &  $^{\circ}$
\end{tabular}
\caption{Predicted quantities}
\end{table}

The CNO Dark Matter contribution to the rotational speeds of a galaxy embedded in the CNO is
\begin{equation}
<v^{2}>|_{R_{C}}(R_{1}) = \frac{GM(R_{1})}{2R_{1}} -  \frac{GM(R_{c})}{2R_{c}}
\end{equation}
where $R_{c}$ is the distance of the galaxy's center from the CNO center, and $R_{1}$ is the distance from the CNO center to the rotational arm of the galaxy. The notation $|_{R_{C}}$ means the centripetal speed is relative to the galaxy's center. The M33 data requires this to be $\sqrt{<v^{2}>|_{R_{C}}}\sim$ 40 km/s at 15 kpc from M33 center, perpendicular to the radial direction. In this equation, $M(R)$ is the CNO enclosed mass at radial coordinate R. It can only be computed numerically\cite{Mor1} by solving the hydro-static equation of equilibrium using the neutrino equation of state. To work out $<v^{2}>$, we use Figure 2.

\begin{figure}
\includegraphics[scale=0.2]{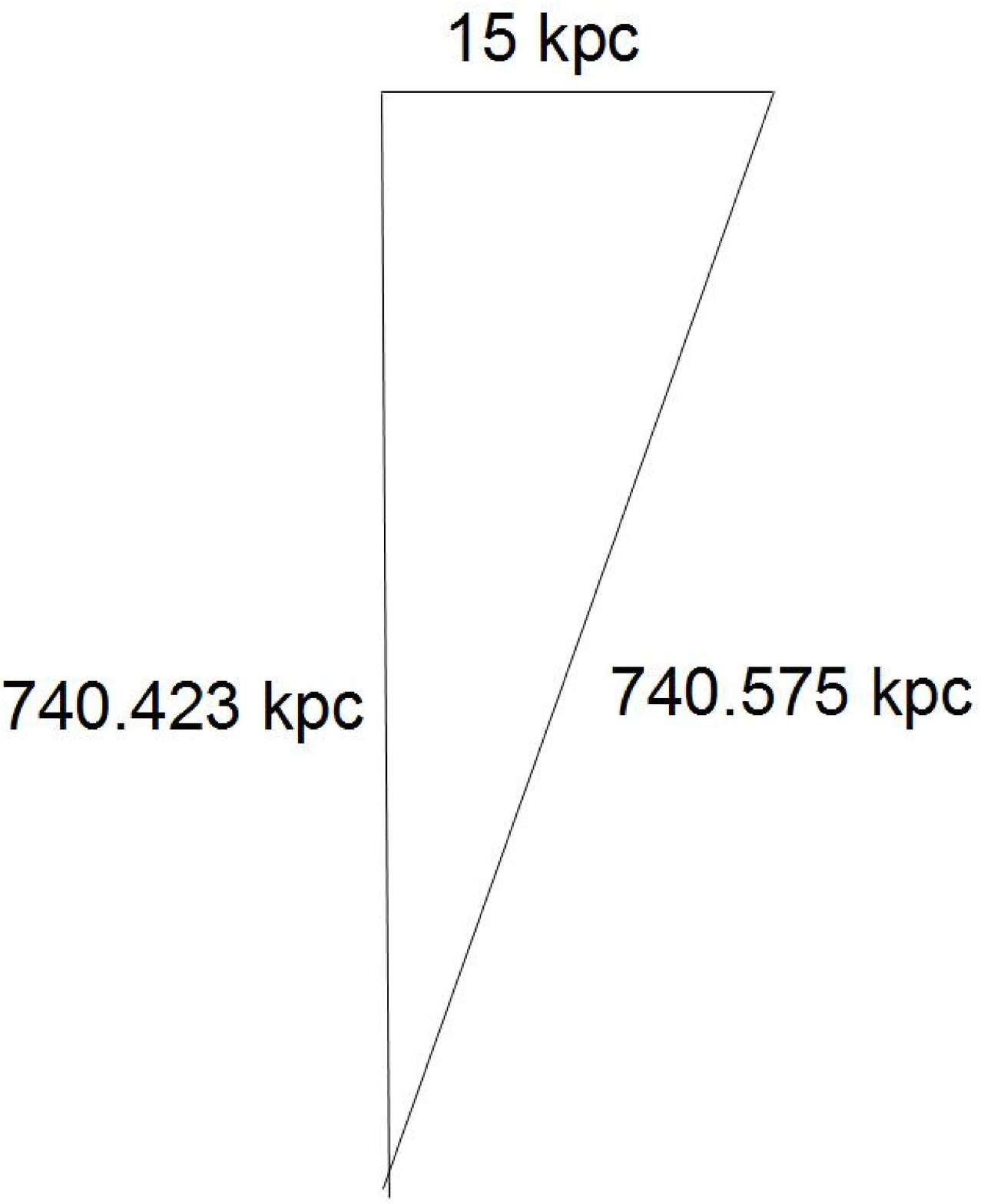}
\caption{Working out the M33 rotation speed at 15 kpc.}
\label{Figure 2}
\end{figure}

In this CNO solution, M33 center is at a radial distance 740.423 kpc, while its 15 kpc spiral arm is at 740.575 kpc. The difference in rotation speeds is given by Eq(5) where $R_{1}$ = 740.575 kpc and $R_{0}$ =  740.423 kpc. By requiring this to be $\sqrt{<v^{2}>|_{R_{C}}}\sim$ 40 km/s at 15 kpc from M33 center, perpendicular to the radial direction, we determine the properties of the CNO embedding the LG. For $x(0) = 0.0225$ with $m \simeq$ 0.8 eV/c$^{2}$, we get the values in Table 4. The value for the Milky Way galaxy is consistent with the highest measured values\cite{OMW}, but they include an unknown mass contribution. The mathematical reason why the Milky Way has a higher rotational speed at 8 kpc than M33 at 15 kpc is due to the non-zero cant angle of Milky Way.  Assuming that the Monte Carlo solution $x(0) = 0.0225$ with $m \simeq$ 0.8 eV/c$^{2}$ is reasonably close to the physical answer, we now can give the mass and radius of this CNO embedding the LG, Table 5.

\begin{table}
    \begin{tabular}{ccc}
\multicolumn{3}{l}{} \\ \hline
Galaxy & spiral arm distance from its center & $<v^{2}>$ \\ \hline
M33 & 15 kpc & $\simeq$ (40 km/s)$^{2}$ \\
Milky Way & 8 kpc &   $\simeq$ (260 km/s)$^{2}$
\end{tabular}
\caption{Dark Matter rotation values for $x(0) = 0.0225$ with $m \simeq$ 0.8 eV/c$^{2}$.}
\end{table}

\begin{table}
    \begin{tabular}{ccc}
\multicolumn{3}{l}{} \\ \hline
quantity & value & unit \\ \hline
CNO-LG boundary condition & $x(0) = 0.0225$ & $\frac{P_{F}}{mc}$ \\
degenerate neutrino mass $m$ & $\simeq 0.80$ & eV/c$^{2}$ \\
CNO-LG radius & 1.316 & Mpc \\
CNO-LG mass & 5.17$\times10^{15}$ & M$_{\odot}$
\end{tabular}
\caption{CNO derived parameters.}
\end{table}

\subsection{LG galaxy motion in the CNO}
A galaxy embedded in a CNO has a gravitational force pulling it in the direction of the CNO center. Eventually the galaxy will undergo simple harmonic motion about the center\cite{Mor1}. Those spiral galaxies that have a non-zero cant angle are also expected to cartwheel about the arm closest to the CNO center. If a cluster of galaxies is embedded, the center of mass of the cluster is itself expected to undergo simple harmonic motion. The back-reaction on the CNO will cause it to vibrate in its spectrum of normal modes\cite{Mor2} with the lowest frequency a quadrupole oscillation. In Figure 3, we show the 3-D geometry of the calculated CNO solution in relationship to the Local Group. It should be pointed out that CNO cannot overlap each other (Pauli Exclusion Principle). Hence, it is possible that CNOs of various radii and masses could be adjacent to one another (like Kepler's stacked greengrocer oranges). In reference\cite{Mor2}, we point out that condensed neutrinos might even form filaments between quasi-spherically shaped CNOs (bi-spherical symmetry). 

\begin{figure}
\includegraphics[scale=0.7]{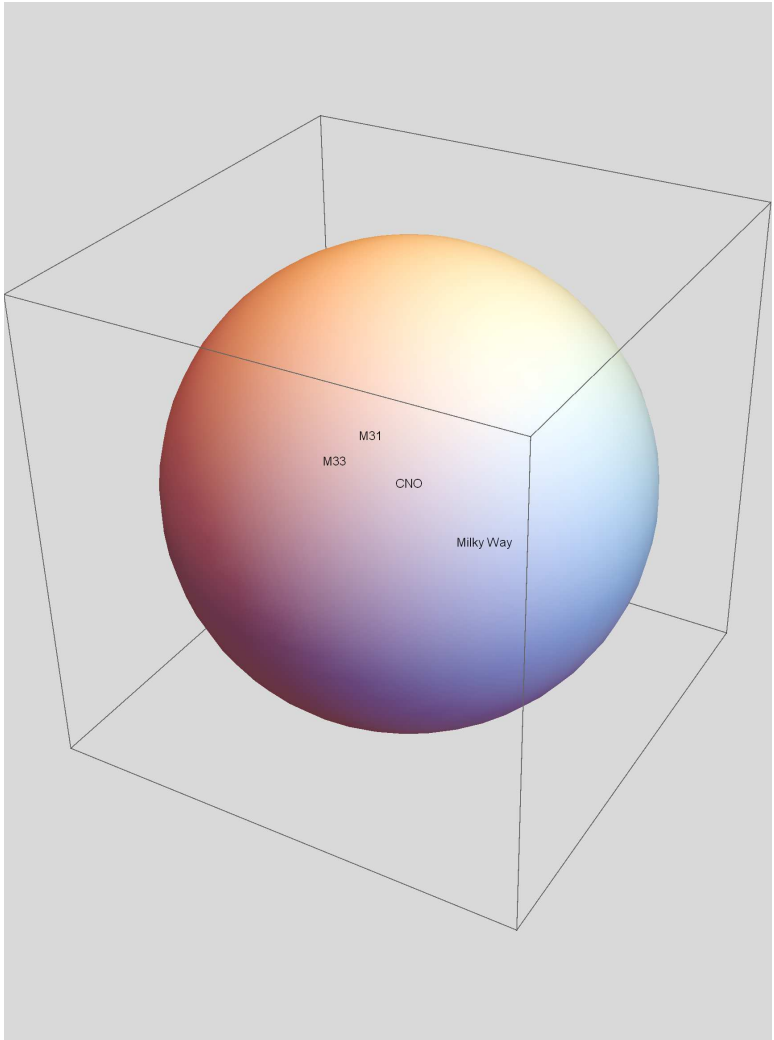}
\caption{Local Group spiral galaxies in the CNO-LG.}
\label{Figure 2}
\end{figure}

\section{Implications for cosmology}
The value of $m \simeq$ 0.8 eV/c$^{2}$ violates the upper-bound limit of neutrinos published by the Planck satellite consortium\cite{PL}, based on their assumption that neutrinos (and anti-neutrinos) are in thermodynamic equilibrium in the early Big Bang, before their decoupling of matter. This thermodynamic equilibrium assumption is conventional wisdom\cite{TU}, which does not take into account possible neutrino (and anti-neutrino) radiation losses. The Planck satellite analysis produces other issues: the discrepancy between the Hubble expansion parameter as indirectly measured by the Planck satellite consortium and the Wilkinson Microwave Anisotropy Probe, with its direct determination by Type 1a supernova\cite{direct}. This tension/disagreement can no longer be ignored because the discrepancy is at the 5\%-9\% level\cite{direct}.  What is going on? The key here is this assumption that the cosmological black-body radiation (CBR) scientists make in reducing their raw data. They assume that the cosmological neutrinos are in thermodynamic equilibrium with baryonic matter right up to their decoupling and that they contibute to the perturbations on the acoustic peaks. In so doing, they constrain neutrino masses to be extremely small and they derive a small value of the Hubble expansion parameter. In fact, if neutrinos radiate their energy in the early universe turbulent plasmas, they will never be in equilibrium with baryonic matter. If so, something else has to be present giving their degrees-of-freedom to the CBR. Actually there is: Adams et al \cite{Adams} show that a cosmic magnetic field has the same contribution to the cosmic microwave background as kinetic degrees of freedom. We anticipate that if the raw CBR data is re-processed using cosmic magnetic fields in place of neutrino ones, the discrepancy of the value of the Hubble parameter should disappear.

Whatever Dark Matter is, it must have an equation of state (EOS). This is the key physics describing the aggregate of Dark Matter particles. A critical requirement is that the EOS allows for stable Dark Matter. Stability in this context means that it must be stable against its own self-gravity. Some internal pressure must exist to withstand its own gravitational collapse. If Dark Matter is made up of bosons, it must have an internal energy source where the kinetic energy of the bosons prevents collapse. No one has shown that candidate bosons have internal energy generation that can prevent the bosons from collapsing into a black hole configuration. Bosons are poor candidates for Dark Matter for this very reason and probably impossible for cold Dark Matter. Fermions, on the other hand, have degeneracy pressure because of the Pauli Exclusion Principle. We know that electron degeneracy pressure allows White Dwarfs to exist, and neutron degeneracy pressure allows Neutron Stars to exist. If neutrinos radiate, they should condense and form stable objects.

If Cold Dark Matter is in fact a condensed object composed of neutrinos, then there are constraints that can be immediately deduced. One constraint is that these neutrinos must be cosmological in origin, since the number of neutrinos produced after the Big Bang is too small. Next, neutrino condensation requires them to loose their kinetic energy. In Eq(1) above, we have a natural way to do this since the early universe had huge turbulent plasmas \cite{Gra, Wag}. However, Majorana neutrinos, have no permanent magnetic dipole moment, and so Eq(1) is not available to them. We conclude that if Dark Matter are objects made of neutrinos, they are of the Dirac-variety. Most importantly, the normal set of neutrinos and their anti-neutrinos ($e, \mu, \tau$) interact so weakly, that whatever numbers were produced in the Big Bang, those numbers still remain and the baryon matter-anti-matter disparity mystery doesn't exist for neutrinos and their antineutrinos. Thus CNO satisfy the large difference between the baryonic contribution (17\%) to the Universe matter density and the Dark Matter contribution (83\%)\cite{DE}.

\section{Experimental consequences}
The KATRIN experiment \cite{KA} will have the sensitivity to determine the mass of the electron anti-neutrino down to $ 0.35 \; eV/c^{2}$, well within the value of $m$ used here. The mass range for the degenerate neutrino mass 0.75 eV/c$^{2} < m < 0.85$ eV/c$^{2}$  found from fitting galaxy cluster data\cite{Mor2} is in direct contradiction to the upper bound claimed by the Planck satellite consortium. If KATRIN discovers a neutrino mass in this range, we contend that the CBR raw data analysis must be revisited and that it would be a major finding endorsing condensed neutrinos as the so-called Dark Matter, which everyone has been looking for.

\clearpage

\section{Appendix}
\appendix
We describe the algorithm that finds the approximate CNO center for two radially-aligned spiral galaxies.
\subsection{Distance of closest approach}
Let $\vec{R_{1}} =\vec{R_{10}} + \vec{a}t$ and $\vec{R_{2}} =\vec{R_{20}} + \vec{b}s$ be the vectors describing the position ($\vec{R_{10}}$) of spiral galaxy numbered 1 and a line through its spin axis ($\vec{a}$), with spiral galaxy numbered 2 similiar. The squared-distance between them $L^{2}(s,t)$ is
\begin{equation}
L^{2}(s,t) = [(\vec{R_{10}}-\vec{R_{20}}) +(\vec{a}t - \vec{b}s)]^{2}
\end{equation}
We find the extremum $s^{\ast}$, $t^{\ast}$ by solving for
\begin{eqnarray}
\frac{\partial L^{2}(s,t)}{\partial s}|_{s = s^{\ast}} & = & 0 \nonumber \\
\frac{\partial L^{2}(s,t)}{\partial t}|_{t = t^{\ast}} & = & 0
\end{eqnarray}
which gives the solutions
\begin{equation}
t^{\ast} = \frac{[(\vec{R_{10}} - \vec{R_{20}})\cdot \vec{a} - (\vec{a} \cdot \vec{b})(\vec{R_{10}} - \vec{R_{20}})\cdot \vec{b}]}{ [(\vec{a} \cdot \vec{b})^{2} -1]}
\end{equation}
\begin{equation}
s^{\ast} = \vec{a} \cdot \vec{b}t^{\ast} + (\vec{R_{10}} - \vec{R_{20}})\cdot \vec{b}
\end{equation}
for unit spin vectors $\vec{a} \cdot \vec{a} = \vec{b} \cdot \vec{b} = 1$.
\subsection{Monte Carlo solution}
We vary the axial ratio and position angle of each spiral galaxy randomly within  $\pm$9 \% of their quoted values within the literature, determine the separation distance between the lines that those probabilty draws create and either keep them as the running smallest value or throw them away before the next probability draw series. A loop of 4 million series draws is sufficient for the approximate answer.

\end{document}